\documentclass[aps,prb,nofootinbib,superscriptaddress,longbibliography]{revtex4-2}
\usepackage{amsmath}
\usepackage{amssymb}
\usepackage{graphicx}
\usepackage{dcolumn}
\usepackage{bm}
\usepackage{comment}
\usepackage{color}
\usepackage{hyperref}
\usepackage{xr-hyper}
\usepackage[english]{babel}
\usepackage[autostyle, english = american]{csquotes}
\usepackage{braket}
\usepackage{pifont}
\MakeOuterQuote{"}

\newcommand{\vect}[1]{\mathrm{\textbf{#1}}}
\newcommand{\op}[1]{\hat{#1}} 
\newcommand{\J}{{J}} 

\begin{document}

\title{Rhombohedral graphite junctions as a platform for continuous tuning between topologically trivial and non-trivial electronic phases}

\author{Luke Soneji}
\affiliation{Department of Physics, University of Bath, Claverton Down, Bath, BA2 7AY, United Kingdom}
\author{Simon Crampin}
\affiliation{Department of Physics, University of Bath, Claverton Down, Bath, BA2 7AY, United Kingdom}
\author{Marcin Mucha-Kruczy\'{n}ski}
\email{M.Mucha-Kruczynski@bath.ac.uk}
\affiliation{Department of Physics, University of Bath, Claverton Down, Bath, BA2 7AY, United Kingdom}


\begin{abstract}
Manipulating the topological properties of quantum states can provide a way to protect them against disorder. However, typically, changing the topology of electronic states in a crystalline material is challenging because their nature is underpinned by chemical composition and lattice symmetry that are difficult to modify. We propose junctions between rhombohedral graphite crystals as a platform that enables smooth transition between topologically trivial and non-trivial regimes distinguished by the absence or presence of topological junction states. By invoking an analogy with the Su-Schrieffer-Heeger model, the appearance of topological states is related to the symmetry of the atomic stacking at the interface between the crystals. The possibility to explore both the topological and non-topological phases is provided by sliding the crystals with respect to each other.
\end{abstract}

\maketitle

\section{Introduction}

Topology, the study of properties invariant under continuous deformations, plays an important role in physics ranging from Gauss's and Ampere's laws of electromagnetism, to the quantum Hall effect \cite{Thouless1982}, optical vortices \cite{Coullet1989} and the properties of space-time \cite{Hawking1972}. The topological properties of quasiparticle dispersions provide a distinct way of categorizing phases of matter \cite{Altland1997}, and underpin the notion of the topological insulator: an important design principle in solid state physics \cite{Hasan2010}, soft matter physics \cite{Tubiana2024}, and photonics \cite{Ozawa2019}. In a topological insulator, an insulating bulk is accompanied by edge states robust to perturbations that do not close the bulk band gap. Arguably the simplest description of edge states is provided by the Su-Schrieffer-Heeger (SSH) model: a one-dimensional chain with alternating strong and weak couplings between nearest neighbour sites which has been extensively used as a framework to describe phenomena in fields as varied as photonics \cite{Schomerus2013, Nathan2022}, excitonics \cite{Davenport2024, Jankowski2025}, plasmonics \cite{Rappoport2021, Xia2023}, magnonics \cite{Li2021, Wei2022}, acoustics \cite{Coutant2021, Li2023}, and circuit electronics \cite{Guo2024, Huang2024}. Originally conceived to study defects/domain walls in polyacetylene \cite{Su1979}, the relative simplicity of the SSH model allows to capture a range of physical effects such as defect dynamics \cite{Su1979}, non-Hermiticity \cite{Schomerus2013} and complex interchain geometries \cite{Sivan2022, Thiang2023}.

Here, we show that rhombohedral graphite junctions provide a realisation of a distinct set of defects in an SSH chain. Atomic stacking and local symmetry at commensurate interfaces between two rhombohedral graphite crystals determine the presence or absence of topological junction states localised at the interface. Furthermore, translating one crystal with respect to the other allows tracking of the evolution of topological junction states as the system transitions between the topologically trivial and non-trivial phases.

\section{Results}

\subsection{Topological junction states at interfaces of rhombohedral crystals}

\begin{figure*}[t]
\centering
\includegraphics[width=\textwidth]{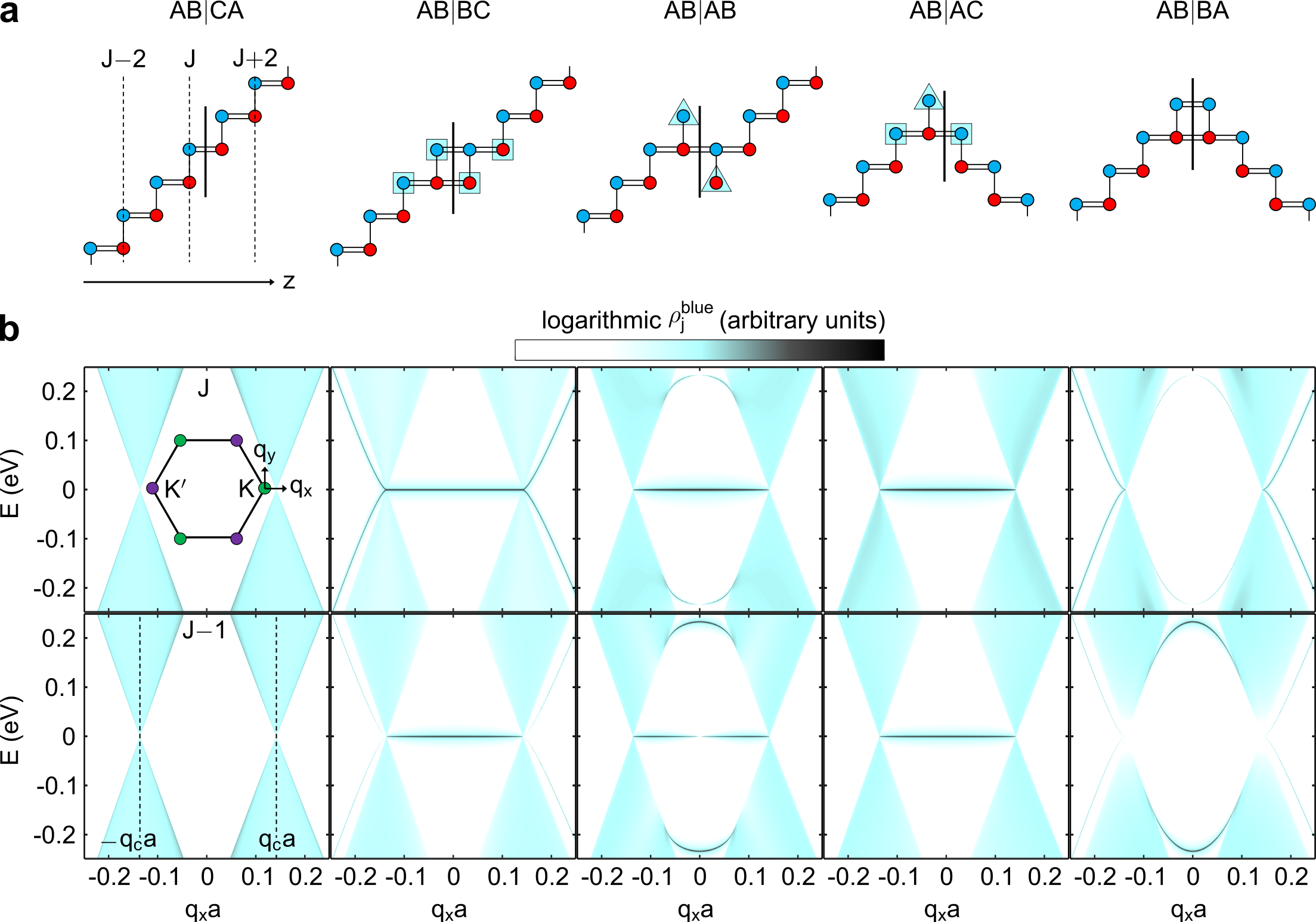}
\caption{\label{fig:topology}
\textbf{Junction geometries and low-energy local density of states.} 
\textbf{a.} Schematic of the five distinct junctions formed by alignment of two rhombohedral graphite half-crystals comprised of layers $\ldots,\J-1,\J$ and layers $\J+1,\J+2,\ldots$ respectively. Layer $\J$ lies directly to the left of the physical interface, indicated with the solid black line. The out-of-plane $z$-axis runs from left to right. Red and blue circles indicate the two inequivalent $A$ and $B$ sublattices in each layer respectively, and the single and double bonds represent intralayer and interlayer coupling between sites respectively. The light blue squares and triangles indicate the sites which host one and two topological states per valley, respectively, at the wave vector $\vect{q}=0$ at the valley centre at the corner of the two-dimensional Brillouin zone. Note that the AB$|$CA, AB$|$BC, and AB$|$AB junctions possess inversion symmetry, while AB$|$AC and AB$|$BA have a mirror reflection plane  which leads to a different sublattice hosting the junction edge state in the right half-crystal. \textbf{b.} The low-energy electronic density of states $\rho^{\mathrm{blue}}_{j}$ on atoms of the blue  ($B$) sublattice in layer $j=\J$ (top row) and $j=\J-1$ (bottom row) as a function of wave vector in the vicinity of the valley $\vect{K}=\left(\tfrac{4\pi}{3a},0\right)$. The in-plane wave vector $\vect{q}=(q_x,q_y)$ is measured from the valley centre (see the inset in the top left panel) and $a$ is the in-plane lattice constant. In these plots, we take $q_{y}=0$. In the bottom left panel we also indicate the critical wave vector $\vect{q}_{c}$ which separates the topologically trivial and non-trivial phases.
}\end{figure*}

Rhombohedral graphite consists of layers of graphene: carbon atoms in a honeycomb arrangement. The layers are regularly stacked such that every atom has a neighbour in one of the adjacent layers, so called ABC stacking equivalent to shifting each layer in succession by one carbon-carbon distance along the bond direction. The real and reciprocal space structures are shown in Supplementary Note 1. We envision bringing together two different rhombohedral graphite crystals into commensurate alignment, and restrict the layer stacking directly at the junction to either rhombohedral (ABC-type), Bernal (ABA-type) or simple-hexagonal (AA-type) alignment \cite{Arovas2008, Taut2013, Taut2014, Taut2016, Sarsfield2025}. The resulting structures are uniquely determined by the stacking of the layers directly at the interface, and information about bulk stacking on both sides (one crystal can have consecutive layers shifted in the same or opposite direction as the other, resulting in ABC or CBA stacking respectively). The latter can be deduced from the arrangement of any two neighbouring layers in each crystal, so that the junctions are fully specified by the four-layer stacking sequence across the interface ($|$), with representative structures AB$|$CA (ideal rhombohedral graphite), AB$|$BC, AB$|$AB, AB$|$AC (equivalent to AB$|$CB), and AB$|$BA. We show these junctions schematically in Fig.\@ \ref{fig:topology}a (where we choose the out-of-plane direction to point from left to right); within each layer atoms occupy one of two distinct sublattices (red and blue).

The alternating nature of the intra- and inter-layer electronic hopping allows mapping of the low-energy electronic structure of rhombohedral graphite onto the SSH model in the limit where only these two couplings are considered \cite{Guinea2006, Manes2007, Muten2021} (we discuss corrections due to higher order couplings in the second half of the paper). Out-of-plane periodicity is preserved in bulk rhombohedral graphite, meaning electronic states possess an out-of-plane wave vector $k$, equivalent to the one-dimensional wave vector in the SSH model. While all junctions but AB$|$CA break this out-of-plane translational symmetry, in-plane periodicity remains and the electronic states possess in-plane wave vector $\vect{q}=(q_{x},q_{y})$ as a good quantum number. This allows a connection between the SSH model and rhombohedral graphite to be made at any fixed $\vect{q}$. Consequently, regardless of the junction geometry, one expects edge states on the outer left and right surfaces of combined crystals. Such states have been observed on the surface of rhombohedral flakes \cite{Koshino2009, Xiao2011, Shi2020, Kaladzhyan2021, Zhang2024}. As we discuss here, the survival and location of the edge states of the two half-crystals at the junction (referred to here as junction states) depend on the atomic stacking present at the interface: AB$|$CA junctions display no junction states; AB$|$BC, AB$|$AB, and AB$|$AC junctions possess eight zero-energy states, while AB$|$BA junctions possess eight dispersing states within the bulk band gap, but at non-zero energy. In Fig.\@ \ref{fig:topology}a, we indicate schematically localization of the zero-energy junction states: for AB$|$BC junctions, these are maximally localised on one of the atomic sites of each of the four junction layers, $\J-1,...,\J+2$; for AB$|$AB junctions, states are maximally localised on the layers $\J$ and $\J+1$; and for AB$|$AC junctions four states are maximally localised on layer $\J$, and two each on layers $\J-1$ and $\J+1$. Note that the two half-crystals are identical and hence possess the same symmetries, meaning that the existence and nature of junction states are prescribed by the coupling at the interface. Junctions AB$|$CA, AB$|$BC and AB$|$AB differ only by an in-plane shift of one half-crystal with respect to the other by one carbon-carbon bond length, as do the twinned systems AB$|$AC and AB$|$BA, yet drastically different junction state configurations result from these atomic-scale changes.

\begin{figure}[t!]
\centering
\includegraphics[width=0.5\textwidth]{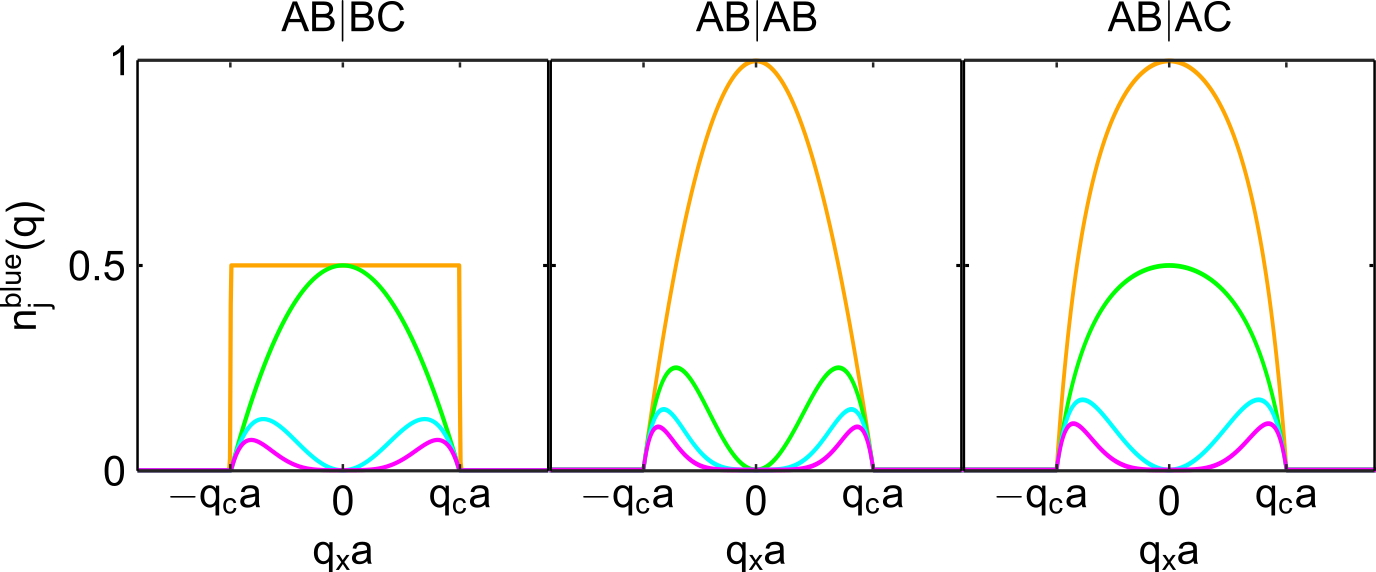}
\caption{\label{fig:decay}
\textbf{Spatial decay of the junction states.} The number of zero-energy states $n_{j}^{\mathrm{blue}}(\vect{q})$ per valley and spin on the blue  ($B$) sublattice atoms of layer $j = \J$, $\J-1$, $\J-2$, $\J-3$ (shown in orange, green, cyan, and magenta, respectively) at the three junctions possessing topological junction states, as a function of wave vector $q_{x}$ for $q_{y}=0$. The corresponding results for the atoms in layers $j > \J$ may be identified by symmetry.
}\end{figure}

As typical for graphene-based systems, the low-energy electronic structure arises in the vicinities of the corners $\vect{K}$ and $\vect{K}^{\prime}$ of the two-dimensional hexagonal Brillouin zone (top-left panel in Fig.\@ \ref{fig:topology}b; see also Supplementary Fig.\@ 1). The panels in Fig.\@ \ref{fig:topology}b show the low-energy electronic dispersion on the blue sublattice atoms in layer $\J$ and $\J-1$ in the vicinity of the "valley" $\vect{K}$ ($\vect{q}=0$) as a function of the dimensionless wave vector $q_{x}a$, with $a$ the in-plane lattice constant (we set $q_{y}=0$; the spectrum in the $\vect{K}^{\prime}$ valley is related to that in $\vect{K}$ by time-inversion). To avoid complications due to finite size effects and the presence of the outer edge states, these results have been calculated assuming semi-infinite half-crystals -- see Methods for more details. For the AB$|$CA junction, which represents the stacking of ideal rhombohedral graphite, the top and bottom panels are identical and show the energy-wave vector regions where bulk states exist. In contrast, the AB$|$BC, AB$|$AB, and AB$|$AC junctions possess dispersionless states at energy $E=0$ that lie outside of this bulk continuum. They occupy the range of wave vector for which the intralayer hopping modulated by the in-plane wave vector is weaker than the interlayer hopping, corresponding to the topologically non-trivial regime of the SSH model \cite{Asbóth2016}. The bulk band gap vanishes at critical wave vector $\vect{q}_{\mathrm{c}}$ (bottom-left panel in Fig.\@ \ref{fig:topology}b), marking the border between topologically trivial and non-trivial regimes. Also like in the SSH model, the topological junction states are entirely localised on only one of the two atomic sites in a given layer, as highlighted in Fig.\@ \ref{fig:topology}a (matching plots of the density of states on the red sublattice atoms are given in Supplementary Note 2). The in-gap dispersive states of the AB$|$AB and AB$|$BA junctions are not  topological; neither are the additional dispersive states existing outside the bulk continuum for $|\vect{q}|>|\vect{q}_{\mathrm{c}}|$ in the AB$|$BC and AB$|$BA junctions. These states are localised at the interface but have non-zero amplitude on atoms of both sublattices.

Close inspection of the top and bottom rows of Fig.\@ \ref{fig:topology}b makes clear that the  topological junction states have an amplitude that depends both upon distance from the junction and on wave vector. These variations are shown more clearly in Fig.\@ \ref{fig:decay} for atoms of the blue sublattice in layers $\J,\ldots,\J-3$. In keeping with the mapping on to a spinless SSH model, for any wave vector $\vect{q}$ in the topological regime there are two zero-energy junction states (one from each half-crystal). At $\vect{q}=0$, these extend over four, two, and three sites at the AB$|$BC, AB$|$AB, and AB$|$AC junctions respectively, becoming increasingly extended as $\vect{q}$ approaches $\vect{q}_{\mathrm{c}}$. Taking into account the electronic spin degeneracy, as well as the electronic states at valley $\vect{K}^{\prime}$, gives the total number of eight junction states in each case. Analytic expressions for the densities of states displayed here are given in Supplementary Note 3, along with the dispersion relations of any topologically trivial junction states.

To explain the presence (or absence) of junction states in each specific system, we focus on the underlying symmetries of the two half-crystals and the junction region. In general terms, the Hamiltonian of each system,
\begin{equation}
\op{H}=\op{H}_{\mathrm{L}} + \op{H}_{\mathrm{R}} + \op{V},
\end{equation}
consists of the left half-crystal described by $\op{H}_{\mathrm{L}}$, right half-crystal described by $\op{H}_{\mathrm{R}}$, and the perturbation $\op{V}$ that couples them together.

The effective one-dimensional real space Hamiltonian of the ABC stacked left half-crystal is
\begin{widetext}
\begin{equation}
\op{H}_{\mathrm{L}} = H_{\mathrm{L},\vect{q}}^{\mathrm{ABC}} = -\gamma_{0}\sum_{j=-\infty}^{\mathrm{J}}\left( f_{\vect{q}}\ket{j}\ket{A}\bra{j}\bra{B}+f_{\vect{q}}^{*}\ket{j}\ket{B}\bra{j}\bra{A} \right) + \gamma_{1}\sum_{j=-\infty}^{\mathrm{J}-1}\left( \ket{j}\ket{B}\bra{j+1}\bra{A} + \ket{j+1}\ket{A}\bra{j}\bra{B}  \right),
\end{equation}
\end{widetext}
where $\gamma_{0}$ and $\gamma_{1}$ are the intra- and interlayer couplings, $f_{\vect{q}}=\exp{\left(\mathrm{i}\tfrac{q_{y}a}{\sqrt{3}}\right)} - \exp{\left(-\mathrm{i}\tfrac{q_{y}a}{2\sqrt{3}}\right)} \left(\cos\tfrac{q_{x}a}{2}+\sqrt{3}\sin\tfrac{q_{x}a}{2}\right)$ is the sum of phase factors due to in-plane nearest neighbours, $\ket{A}$ and $\ket{B}$ are the in-plane sublattice Bloch states  on the red ($A$) and blue ($B$) sites, and $\ket{j}$ captures the layer degree of freedom with $j$ indexing the layers. The Hamiltonian of an ABC stacked right half-crystal $\op{H}_{\mathrm{R}} = H_{\mathrm{R},\vect{q}}^{\mathrm{ABC}}$ takes the same form, with $j \geq \J+1$. If the right half-crystal is instead CBA stacked, one needs to swap the $A$ and $B$ sublattices in the interlayer coupling term. Finally, the junction perturbation $\op{V}$ takes one of three forms prescribed by the relative stacking of layers $\J$ and $\J+1$:  $\op{V} = \op{V}^{\mathrm{B|C}}$ for the AB$|$CA junction, $\op{V} = \op{V}^{\mathrm{B|A}}$ for AB$|$AB and AB$|$AC, and $\op{V} = \op{V}^{\mathrm{B|B}}$ for AB$|$BC and AB$|$BA, with
\begin{align}\begin{split} \label{eqn:perturbation}
&\op{V}^{\mathrm{B|C}} = \gamma_{1}\ket{\J}\ket{B}\bra{\J+1}\bra{A}+\mathrm{H.c.}, \\
&\op{V}^{\mathrm{B|A}} = \gamma_{1}\ket{\J}\ket{A}\bra{\J+1}\bra{B}+\mathrm{H.c.}, \\
&\op{V}^{\mathrm{B|B}} = \gamma_{1}\ket{\J}\bra{\J+1}(\ket{A}\bra{A}+\ket{B}\bra{B})+\mathrm{H.c.}.
\end{split}\end{align}

 In the absence of perturbation, both half-crystals host zero-energy edge states, $\ket{\alpha_{\mu}}$, where $\mu=l,r$ denotes the left ($l$) or right ($r$) edge of the left ($\alpha=\mathrm{L}$) or right ($\alpha=\mathrm{R}$) half-crystal. Because $\op{V}$ is local to the interface, its matrix elements involving both the bulk and edge states of the half-crystals must vanish in the thermodynamic limit. As a result, the energies of the edge states at the interface, $\ket{\mathrm{L}_{r}}$ and $\ket{\mathrm{R}_{l}}$, are only modified if they are directly coupled by the perturbation, i.\@e.\@ $\braket{\mathrm{L}_{r}|\op{V}|\mathrm{R}_{l}} \neq 0$. Each edge state has support on only one sublattice, which is specified by the final site at the respective edge. This means that the condition necessary to remove the topological edge states at the interface is equivalent to requiring dimerization, $\bra{\J}\braket{r|\op{V}|\J+1}\ket{l}\neq 0$, with $r$ and $l$ now denoting the sublattices hosting the edge states at the interface. Note that, while for the left half-crystal we always have $r=B$, we have $l=A$ for ABC stacked right half-crystal but $l=B$ for CBA stacked one (this is equivalent to distinguishing junctions with inversion symmetry: AB$|$CA, AB$|$BC and AB$|$AB, from those with mirror reflection: AB$|$AC and AB$|$BA). From Eq.\@ \ref{eqn:perturbation}, it follows that the topological zero-energy states persist for the AB$|$BC, AB$|$AB and AB$|$AC junctions, while they hybridize and hence gap out for the AB$|$CA and AB$|$BA junctions.

\subsection{Topological transitions in sliding crystals}

\begin{figure*}[t]
\centering
\includegraphics[width=1.0\textwidth]{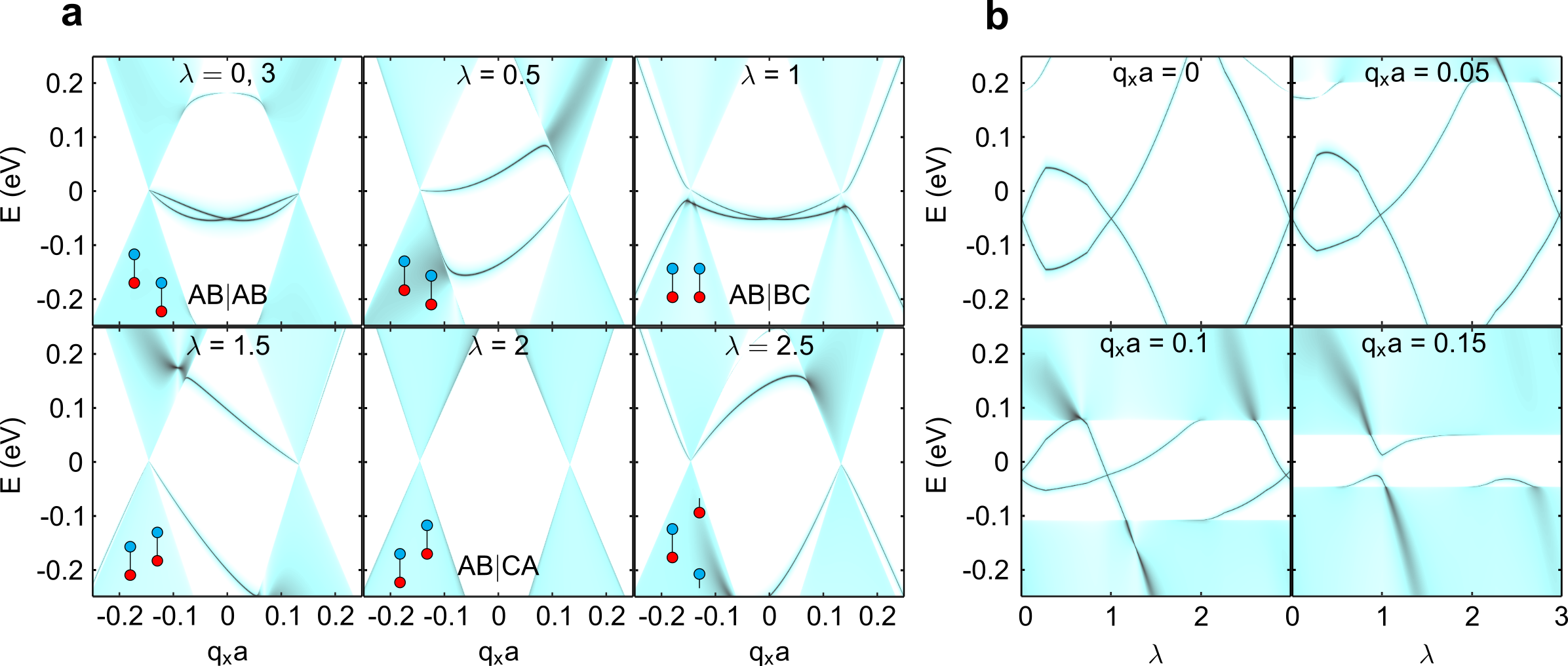}
\caption{\label{fig:sliding}
\textbf{Evolution of the topological states for sliding of the half-crystals.} \textbf{a.} The wave-vector resolved electronic density of states $\rho^{\mathrm{blue}}_{\J}$ on the blue sublattice of layer $\J$ at fixed points during the sliding process as a function of the dimensionless parameter $\lambda$, starting from the AB$|$AB junction ($\lambda=0$ and also $\lambda=3$ as the sliding structure is periodic), through the AB$|$BC ($\lambda=1$) to the AB$|$CA junction ($\lambda=2$). \textbf{b.} Evolution of the electronic states at various wave vectors as a function of $\lambda$.
}\end{figure*}

The three junctions AB$|$CA, AB$|$BC and AB$|$AB are related to one another by in-plane translational shifts of one carbon-carbon bond-length; the AB$|$AC and AB$|$BA junctions are similarly related. This enables the possibility of moving continuously between the topologically trivial and non-trivial phases in a single system, by sliding one rhombohedral graphite crystal with respect to the other in the plane of the interface. Describing such sliding necessitates going beyond the nearest-neighbour SSH model, in order to account for changes of electronic couplings as the half-crystals are moved away from the junction configurations considered so far and the distances between atomic sites at the interface change. To explore the qualitative effects we adopt a simple distant-dependent interlayer hopping (see Methods and Supplementary Note  4 for more details). These interactions are included at the junction and throughout both half-crystals, effectively leading to the inclusion of skew interlayer hoppings in our description of rhombohedral graphite \cite{Dresselhaus1981}. Finally, we parametrize the sliding configurations with dimensionless $\lambda$ such that $\lambda a$ is the in-plane translation of the right half-crystal in the bond direction.

Fig.\@ \ref{fig:sliding}a shows the evolution of the low-energy density of states starting from AB$|$AB ($\lambda=0$), through to AB$|$BC ($\lambda=1$), AB$|$CA ($\lambda=2$) and back to AB$|$AB ($\lambda=3$). Additional results for intermediate values of $\lambda$ are given in Supplementary Note  5, along with results for the sliding process from AB$|$AC to AB$|$BA. Comparison of the top-left panel with Fig.\@ \ref{fig:topology}b shows that inclusion of further interlayer couplings  shifts the energy of the junction states at $\vect{q}=0$ away from zero, lifts their degeneracy at $\vect{q} \neq 0$, and introduces dispersion. This arises because the atomic sites hosting the junction states are now directly coupled. However, all interlayer couplings other than the direct vertical one introduce terms that are modulated by the in-plane wave vector via $f_{\vect{q}}$ so that the interface states remain degenerate at the valley centre, $\vect{q}=0$, where $f_{\vect{q}=0}=0$. Degeneracy is also observed at $|\vect{q}|=|\vect{q}_{\mathrm{c}}|$, where the impact of dimerization drops to zero because, for the AB$|$AB junction, the amplitude of zero-energy states on the edge sites vanishes as $\vect{q} \rightarrow \vect{q}_{\mathrm{c}}$ (see Fig.\@ \ref{fig:decay}).  The energy of the junction states at $\vect{q}=0$ is no longer zero because the new hoppings break electron-hole symmetry \cite{Marcin2010} and we take energy $E=0$ to be the Fermi level of the bulk crystal which must lie at the touching point of the continua. Note the symmetry-breaking hoppings are small in magnitude compared to the band gap, so their influence does not cause the removal of the topological junction states from within the gap.

As the right crystal is displaced from the AB$|$AB configuration which hosts topological states (top-left panel in Fig.\@ \ref{fig:sliding}a) and $\lambda$ is increased from zero, breaking of the $C_{3}$ symmetry at the interface means that previously equivalent hoppings become different and non-zero dimerization occurs even at $\vect{q}=0$, leading to splitting of the junction states. As seen in Fig.\@ \ref{fig:sliding}b which shows the evolution with $\lambda$ of the electronic states at a fixed wave vector, the energy splitting increases to a maximum before decreasing so that the states again become degenerate at the valley centre for $\lambda=1$, the AB$|$BC configuration. For this configuration, the junction states are not degenerate at $|\vect{q}|=|\vect{q}_{\mathrm{c}}|$; this is due to non-zero dimerization of the edge sites at all $\vect{q} \neq 0$. Instead, the junction state on the valence band side overlaps with the continuum. As the right crystal is slid further, the energy  splitting of the junction states again increases until the states merge entirely with the continuum when the AB$|$CA junction configuration is reached at $\lambda=2$. On further sliding discrete states emerge again, eventually returning to degeneracy at $\vect{q}=0$ when the AB$|$AB junction is reformed at $\lambda=3$. The repeated disappearance and emergence of the junction states from the bulk continuum as the crystals are displaced with respect to each other can be seen in Fig.\@ \ref{fig:sliding}b, especially for wave vectors away from $\vect{q}=0$. Also shown is an example of the evolution with $\lambda$ of the low-energy dispersion for a wave vector outside the topologically non-trivial range, $q_{x}a=0.15>q_{\mathrm{c}}a$.  The states intermittently splitting from the continua do not originate in the zero-energy edge states of the half-crystals and are not topological.

\section{Discussion}

\begin{figure*}[t]
\centering
\includegraphics[width=\textwidth]{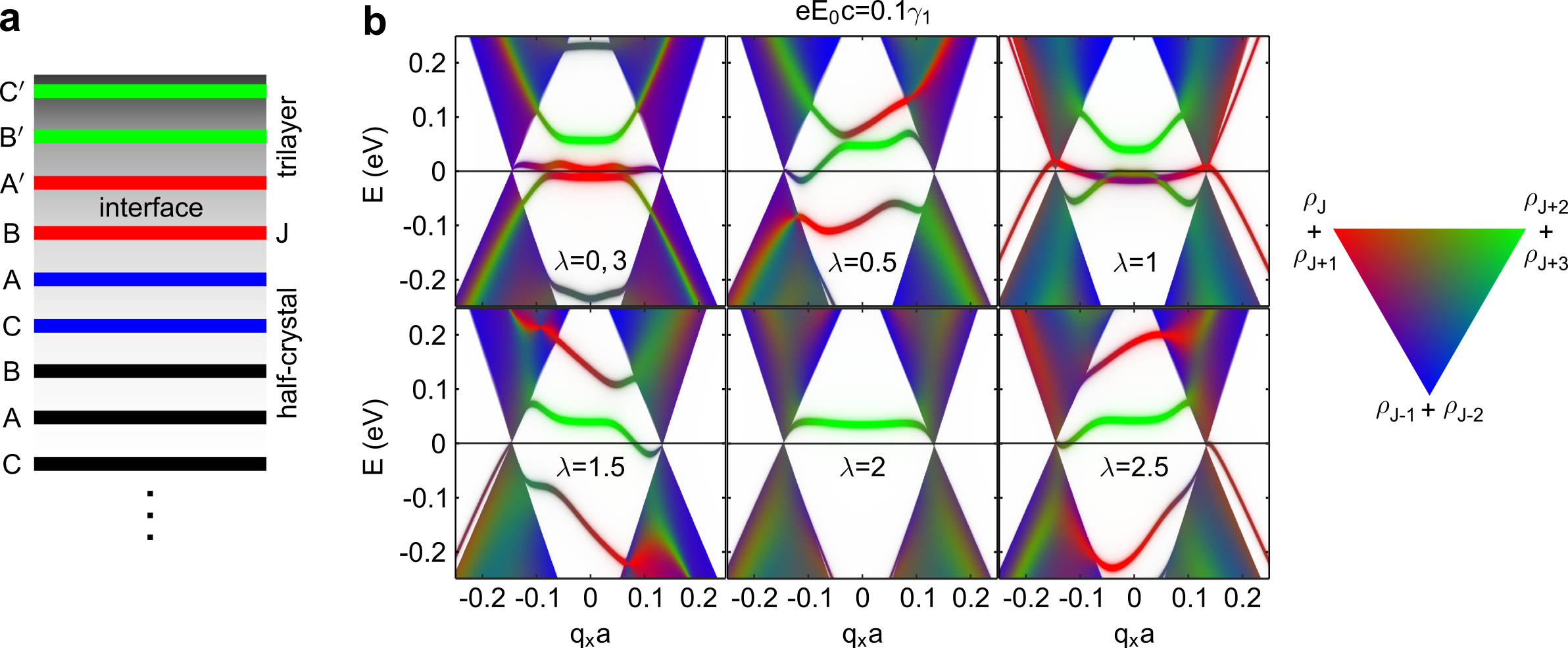}
\caption{\label{fig:gating}
\textbf{Junction states in rhombohedral trilayer on a rhombohedral half-crystal with applied electric field.} \textbf{a.} Visualisation of the heterostructure. The layers are colour coded as a key to panel b, and the grey shading represents strength of the electric field screened by the layers of the crystal. The trilayer is ABC stacked when considered in isolation, with the overall stacking sequence of the heterostructure determined by the stacking configuration at the interface. \textbf{b.} The low-energy electronic density of states $\rho_{j}$ for rhombohedral trilayer graphene on top of a rhombohedral half-crystal with the interface stacking parametrized by the sliding parameter $\lambda$ like in Fig.\@ \ref{fig:sliding}. An out-of-plane electric field $E_{0}$ is applied to the heterostructure such that $eE_{0}c=0.1\gamma_{1}$. The amount of green, red and blue colour reflects the fraction of the electronic states localised on the two surface layers ($\J+2$ and $\J+3$) of the heterostructure, the two layers either side of the physical interface ($\J$ and $\J+1$), and the $\J-1$ and $\J-2$ layers of the heterostructure respectively. The transparency of the colour at each point varies linearly, scaled by the total density of states across these six layers of the heterostructure.
}\end{figure*}

We have shown that the atomic-scale geometry determines the topological properties and the presence of junction states at rhombohedral graphite junctions. While it could be argued that no bonds are explicitly broken when sliding two crystals with respect to one other, such a process alters the hierarchy of dominant interlayer couplings and effectively involves the tearing and reforming of bonds. Despite changes at the level of a single inter-atomic bond, this does not constitute a continuous deformation that leaves the topology of the system unchanged and provides a cautionary example of how atomistic details might modify any "topologically robust" properties. Crucially, the rhombohedral graphite system enables investigations of both topological and non-topological phases, and the transition between them.

To exclusively focus on junction properties without complications due to other boundaries our calculations have considered semi-infinite half-crystals rather than finite-thickness films. For experimental realizations, it is necessary to consider how thick the rhombohedral crystals must be to observe the junction states. The primary impact of finite thickness is to discretize the continuum states without affecting the topological properties of the junction. For Bernal-stacked graphene it is typically assumed that films of ten or more layers behave electronically like graphite \cite{Partoens2006}. Applied to rhombohedral crystals, this would imply compound crystals of $\sim20$ layers. It has been shown \cite{Aitor2019, Palinkas2024} that electronic Raman scattering can distinguish between Bernal and rhombohedral stacked crystals, and that stacking faults like those discussed here lead to observable features \cite{Aitor2019} in Raman spectra, enabling an indirect measure of junction electronic structure.

For thinner films, it would be possible to use surface sensitive methods as applied to two-dimensional crystals like graphene multilayers \cite{Liu2018} to probe the junction. One complication is that the signal from the surface state on the outer surface of a heterostructure made of rhombohedral crystals might overshadow signatures of the interface state. However, we suggest that this can be avoided by applying an out-of-plane electric field using an electrostatic gate \cite{Taychatanapat2010, Onodera2019}, through surface doping \cite{Kang2017, Zhang2024, Lim2024}, or tunnelling tips which induce local fields \cite{Ariskina2021}; this would split the energies of the surface and junction states. Fig.\@ \ref{fig:gating}a shows an example of a heterostructure consisting of a rhombohedral trilayer placed upon a rhombohedral-half crystal. The wave vector-resolved electronic density of states for the external field $E_{0}=0.1\tfrac{\gamma_{1}}{ec}\approx10^{-2}$ V \AA$^{-1}$ \ is shown in Fig.\@ \ref{fig:gating}b--see Supplementary Note 6 for the results with no external field. We have determined the impact of this electric field by performing a self-consistent calculation of charge redistribution amongst all the layers in the heterostructure \cite{Koshino2010}. We use green, red, and blue to show the fraction of the electronic state at a given wave vector localised on the two surface layers ($\J+2$ and $\J+3$) of the heterostructure, the two layers either side of the physical interface ($\J$ and $\J+1$), and the $\J-1$ and $\J-2$ layers respectively. Because of the finite thickness of trilayer graphene, dispersion of the interface states differs in details as compared to Fig.\@ \ref{fig:sliding}a. Nevertheless, the interface states are well separated in energy from the surface state, and follow a similar evolution with $\lambda$ as seen before. For the AB$|$AB junction, $\lambda=0$, the green state on the outer surface is clearly distinguished from the red interface states. One of the interface states for the AB$|$BC junction, $\lambda=1$, is green, because even at the valley centre it is localised on layers $\J-1$ and $\J+2$. This demonstrates the difference between the topologically non-trivial configurations AB$|$AB and AB$|$BC.

Given the tremendous progress in graphene-based van der Waals heterostructure fabrication \cite{Yang2019, Lau2022} and their mechanical manipulation \cite{RibieroPalau2018, Hu2022, Yeo2025}, as well as availability of rhombohedral crystals with thickness from a few to tens of layers \cite{Shi2020, Kaladzhyan2021, Zhang2024}, we anticipate that the fascinating and richly structured junction states proposed can be realised and explored experimentally. Different junctions could be probed either explicitly by moving one of the crystals or by investigating differently stacked local domains. For the former, for a constant sliding force, the change in surface friction as a function of $\lambda$ might cause the sliding velocity to vary. We note that, while AA stacking is energetically unfavourable, the reflection symmetry with respect to the sliding direction prevents the crystals from avoiding this configuration by following a different route. Hence, we believe it should be possible to explore stackings close to the AB$|$BC and AB$|$BA configurations. For the latter approach of exploring different domains in a static experiment, introducing a small twist between the crystallographic directions of the two half crystals will lead to the formation of local AA, AB, and BA domains at the junction, similar to the case of graphite and graphene moiré superlattices \cite{Rong1993, Brihuega2012}.

While our general approach to junctions of semi-infinite half crystals can be applied to other layered materials, the link between edge states and the SSH model is unique to rhombohedral graphite. About $15\%$ of all naturally occurring graphite is rhombohedral stacked, with the vast majority taking the ABA Bernal form \cite{Haering1958, Lui2011}. For that reason, it would be interesting to investigate junctions of half-crystals with different stackings, with rhombohedral regions serving as a potential source of topological states.

\section{Methods}

\subsection{Calculating the density of states}

We calculate our (layer- or sublattice-resolved if necessary) electronic density of states results by embedding the four-layer junction region between two semi-infinite rhombohedral half-crystal surfaces. This is done via calculation of the junction Green's function $G^{\J}$:
\begin{equation}\label{eqn:Gembed}
G^{\J} (z, \vect{q}) = \left[ z - H^{\J} (\vect{q}) - \Sigma^{\mathrm{L}} (z, \vect{q}) - \Sigma^{\mathrm{R}} (z, \vect{q}) \right]^{-1},
\end{equation}
where  here, $z$ is the complex energy parameter, $H^{\J}$ is the 8$\times$8 junction Hamiltonian describing only layers $\J-1$ to $\J+2$, and $\Sigma^{\mathrm{L}}$ and $\Sigma^{\mathrm{R}}$ are embedding potentials accounting exactly for the influence of the rhombohedral half-crystals either side of the four-layer junction region. We use the Slonczewski-Weiss-McClure tight-binding model \cite{Dresselhaus1981} to construct $H^{\mathrm{J}}$, but for Fig.\@ \ref{fig:topology} of the main text limit ourselves to the in-plane nearest neighbour hopping $\gamma_{0}$ and the vertical interlayer hopping $\gamma_{1}$. Labelling the red sites as sublattice $A$ and the blue sites as sublattice $B$, the general form in the basis of the Bloch states on sublattices $A_{\J-1}$, $B_{\J-1}$, $A_{\J}$, $B_{\J}$, $A_{\J+1}$, $B_{\J+1}$, $A_{\J+2}$, $B_{\J+2}$, is
\begin{equation} \label{eqn:junctionH}
H^{\J}(\vect{q}) = 
\begin{pmatrix}
H_{0} & V_{\J-1,\J} & 0 & 0 \\
V_{\J-1,\J}^{\dagger} & H_{0} & V_{\J,\J+1} & 0 \\
0 & V_{\J,\J+1}^{\dagger} & H_{0} & V_{\J+1,\J+2} \\
0 & 0 & V_{\J+1,\J+2}^{\dagger} & H_{0}
\end{pmatrix}.
\end{equation}
The diagonal blocks are the graphene monolayer Hamiltonians,
\begin{equation} \label{eqn:monolayerH}
H_{0}(\vect{q}) = 
\begin{pmatrix}
0 & -\gamma_{0} f_{\vect{q}}\\
-\gamma_{0}f^{*}_{\vect{q}} & 0
\end{pmatrix}.
\end{equation}
The off-diagonal blocks $V_{j,j+1}$ describe interlayer coupling between neighbouring layers $j$ and $j+1$. Here, $V_{\J-1,\J} = V_{\mathrm{AB}} = \tfrac{1}{2}\gamma_{1}(\sigma_{x}-\mathrm{i}\sigma_{y})$, with $\sigma_{x}$ and $\sigma_{y}$ the $x$ and $y$ Pauli matrices. For the AB$|$CA, AB$|$BC, and AB$|$AB junctions, $V_{\J+1,\J+2} = V_{\mathrm{AB}}$, and for the AB$|$BA and AB$|$AC junctions, $V_{\J+1,\J+2} = V_{\mathrm{AB}}^{T}$. The coupling matrix $V_{\J,\J+1}$ depends on the configuration of the layers at the junction; for the AB$|$CA junction $V_{\J,\J+1}=V_{\mathrm{AB}}$, for AB$|$BC and AB$|$BA $V_{\J,\J+1}=\gamma_{1}\sigma_{0}$, with $\sigma_{0}$ the $2\times 2$ identity matrix, and for AB$|$AB and AB$|$AC $V_{\J,\J+1}=V_{\mathrm{AB}}^{T}$. We use the values $\gamma_{0}=3.16$\ eV and $\gamma_{1}=0.38$\ eV \cite{Slizovskiy2019} and the in-plane lattice parameter $a = 2.46$ \AA \ \cite{Dresselhaus1981}.

In the calculation of $G^{\J}$, only the $j,j'=\J-1,\J-1$ block of $\Sigma^{\mathrm{L}}$ and the $j,j'=\J+2,\J+2$ block of $\Sigma^{\mathrm{R}}$ are non-zero. They are explicitly,
\begin{subequations}
\begin{equation}
\Sigma^{\mathrm{L}}_{\J-1,\J-1} (z, \vect{q}) = V_{\mathrm{AB}}^{\dagger} G_{\J-2,\J-2}^{\mathrm{L}} (z, \vect{q}) V_{\mathrm{AB}},
\end{equation}
\begin{equation}
\Sigma^{\mathrm{R}}_{\J+2,\J+2} (z, \vect{q}) = V_{\J+2,\J+3} G_{\J+3,\J+3}^{\mathrm{R}} (z, \vect{q}) V_{\J+2,\J+3}^{\dagger}.
\end{equation}
\end{subequations}
Here, $G_{\J-2,\J-2}^{\mathrm{L}}$ and $G_{\J+3,\J+3}^{\mathrm{R}}$ are the $2\times 2$ surface blocks of the Green's function for the isolated rhombohedral half-crystals on the left and right. We derive an analytic expression for $G^{\mathrm{L}}$ from a modified Eqn.\@ \ref{eqn:Gembed} where $H^{\J} = H_{0}$ and $\Sigma^{\mathrm{R}}=0$:
\begin{equation}\label{eqn:GLeqn}
G^{\mathrm{L}} (z, \vect{q}) = \left[ z - H_{0} (\vect{q}) - V_{\mathrm{AB}}^{\dagger} G^{\mathrm{L}} (z, \vect{q}) V_{\mathrm{AB}} \right]^{-1},
\end{equation}
with solution
\begin{equation}\label{eqn:GLresult}
G^{\mathrm{L}} (z, \vect{q}) = \frac{\beta}{z - \gamma_{1}^{2} \beta}
\begin{pmatrix}
z & -\gamma_{0}f_{\vect{q}} \\
-\gamma_{0}f^{*}_{\vect{q}} & z - \gamma_{1}^{2} \beta
\end{pmatrix},
\end{equation}
where
\begin{equation}\label{eqn:beta}
\beta = \frac{z^{2} + \gamma_{1}^{2} - \gamma_{0}^{2} |f_{\vect{q}}|^{2} + \sqrt{\left( z^{2} + \gamma_{1}^{2} - \gamma_{0}^{2} |f_{\vect{q}}|^{2} \right)^{2} - 4 \gamma_{1}^{2} z^{2}}}{2 \gamma_{1}^{2} z}.
\end{equation}
To obtain $G^{\mathrm{R}}$ for an ABC stacked right half-crystal we interchange the diagonal elements of $G^{\mathrm{L}}$. If the right-half crystal is CBA stacked, $G^{\mathrm{R}}=G^{\mathrm{L}}$.

The local density of states $\rho(E, \vect{q})$ on site $\mu$ of layer $j$ is then
\begin{equation} \label{eqn:ldos}
\rho^{\mu}_{j}(E,\vect{q})= - \frac{1}{\pi} \mathrm{Im} \left[ G^{\mu,\mu}_{j,j} (E + \mathrm{i} \eta, \vect{q}) \right],
\end{equation}
where $\eta$ circumvents difficulties due to poles but results in a $2\eta$ full width at half maximum Lorentzian broadening of spectral features. We take $\eta \leq 0.1$ meV in this paper.

\subsection{Modelling a sliding junction}

To model a sliding junction, we allow all sites to couple to sites in the neighbouring layers and introduce distance dependence into the interlayer coupling. This effectively introduces interlayer couplings $\gamma_{3}$ and $\gamma_{4}$ from the Slonczewski-Weiss-McClure model of graphite; however, our assumption that the interlayer hopping depends only the distance between sites means we do not distinguish explicitly between these two couplings. We adopt a Slater-Koster scheme \cite{Slater1954} in which a hopping between a $p_{\mathrm{z}}$ orbital on the atomic site $\mu$ on layer $\J$ and a $p_{\mathrm{z}}$ orbital on the site $\mu'$ on layer $\J+1$ and distance $d$ away is,
\begin{equation}
\gamma^{\mu,\mu'} = V_{\mathrm{pp}\sigma} (d^{\mu,\mu'}) \cos^{2} (\theta^{\mu,\mu'}),
\end{equation}
where $V_{\mathrm{\mathrm{pp}}\sigma}(d^{\mu,\mu'})$ is the distance-dependent $\sigma$-bond integral, and $\theta^{\mu\mu'}$ is the angle between the $z$-axis and the vector connecting the sites $\mu$ and $\mu'$. Here we take $c=3.35$ \AA \ as the interlayer distance \cite{Dresselhaus1981}. Note that we neglect the contribution from the $\pi$-bond integral $V_{\mathrm{pp}\pi}$ as the distances involved are comparable to next-nearest in-plane neighbour distances, and we do not include the corresponding in-plane couplings in our description. An interlayer matrix element of the Hamiltonian includes a sum over all sites of the same type for which $V_{\mathrm{pp}\sigma} (d)>0$, with each term weighted by an appropriate phase factor. Further details of the model are given in Supplementary Note 4. Note that, for consistency, we also include the skew interlayer hopping between layers across the rhombohedral half-crystals (in this case, only $d=\sqrt{c^{2}+\left(\tfrac{a}{\sqrt{3}}\right)^2}$ is relevant). Due to the increased number of non-zero terms in the respective Hamiltonians, $\Sigma^{\mathrm{L}}$ and $\Sigma^{\mathrm{R}}$ are now obtained numerically by decimation \cite{LannooFriedel}.

\section*{Acknowledgments}
This work has been supported by the UK Engineering and Physical Sciences Research Council (EPSRC) through the Grant EP/W524712/1.

\section*{Data availability}
No datasets were generated or analysed during the current study. All information presented in the figures of the paper can be obtained from the equations presented therein or in the supplementary material.

\section*{Author contributions}
M.\@M.\@-K.\@ conceived the project. L.\@S.\@ carried out the theoretical calculations and analysis with the assistance of S.\@C.\@ and M.\@M.\@-K. All authors built the theoretical model and wrote the manuscript.

\section*{Competing interests}
The authors declare no competing interests.

\section*{References}
\bibliography{References.bib}

\end{document}


\title{Supplementary Information: \\ Rhombohedral graphite junctions as a platform for continuous tuning between
topologically trivial and non-trivial electronic phases}

\author{Luke Soneji}
\affiliation{Department of Physics, University of Bath, Claverton Down, Bath, BA2 7AY, United Kingdom}
\author{Simon Crampin}
\affiliation{Department of Physics, University of Bath, Claverton Down, Bath, BA2 7AY, United Kingdom}
\author{Marcin Mucha-Kruczy\'{n}ski}
\email{M.Mucha-Kruczynski@bath.ac.uk}
\affiliation{Department of Physics, University of Bath, Claverton Down, Bath, BA2 7AY, United Kingdom}

\date{\today}

\maketitle

\tableofcontents
\newpage

\section*{Supplementary Note 1: Real and reciprocal lattice geometry}

Rhombohedral graphite has a unit cell with primitive lattice vectors $\vect{a}_1 = a(-1/2,\sqrt{3}/2,0)$,
$\vect{a}_2 = a(1/2,\sqrt{3}/2,0)$, and
$\vect{a}_3 = (0,a/\sqrt{3},c)$. The in-plane lattice parameter is $a = 2.46$ \AA \ and the interlayer spacing is $c = 3.35$ \AA \ \cite{Dresselhaus1981}. Because of broken out-of-plane periodicity for all junctions except AB$|$CA, only the in-plane wave vector is a good quantum number, and the relevant Brillouin zone is that of two-dimensional graphene.

\begin{figure}[h!]
\centering
\includegraphics[width=0.7\textwidth]{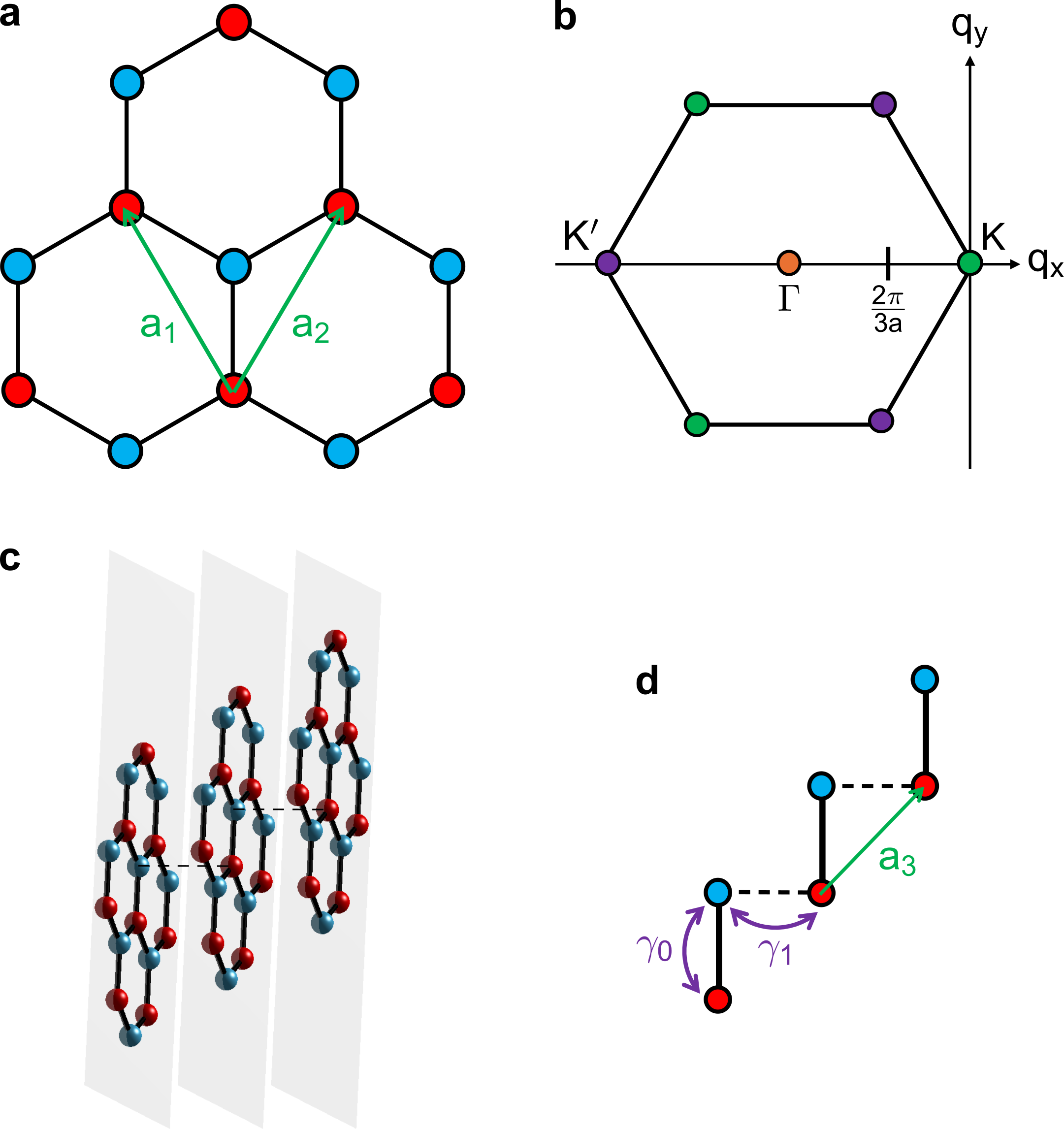}
\caption{\label{fig:structure} 
\textbf{Real and reciprocal lattice geometry.} \textbf{a.} In-plane real space structure of a graphene monolayer.  Red (blue) circles indicate the positions of atoms belonging to  two different sublattices. \textbf{b.} The hexagonal Brillouin zone of graphene. Equivalent corners are denoted with like colours. \textbf{c.} A trilayer section of rhombohedral stacked graphite. \textbf{d.} Side view of the trilayer section. Hopping processes included in the Hamiltonian used in the minimal model are denoted by $\gamma_0$ and $\gamma_1$.
}
\end{figure}

\pagebreak

\section*{Supplementary Note 2: Density of states on red sublattice atoms}

\begin{figure}[h!]
\centering
\includegraphics[width=1.00\textwidth]{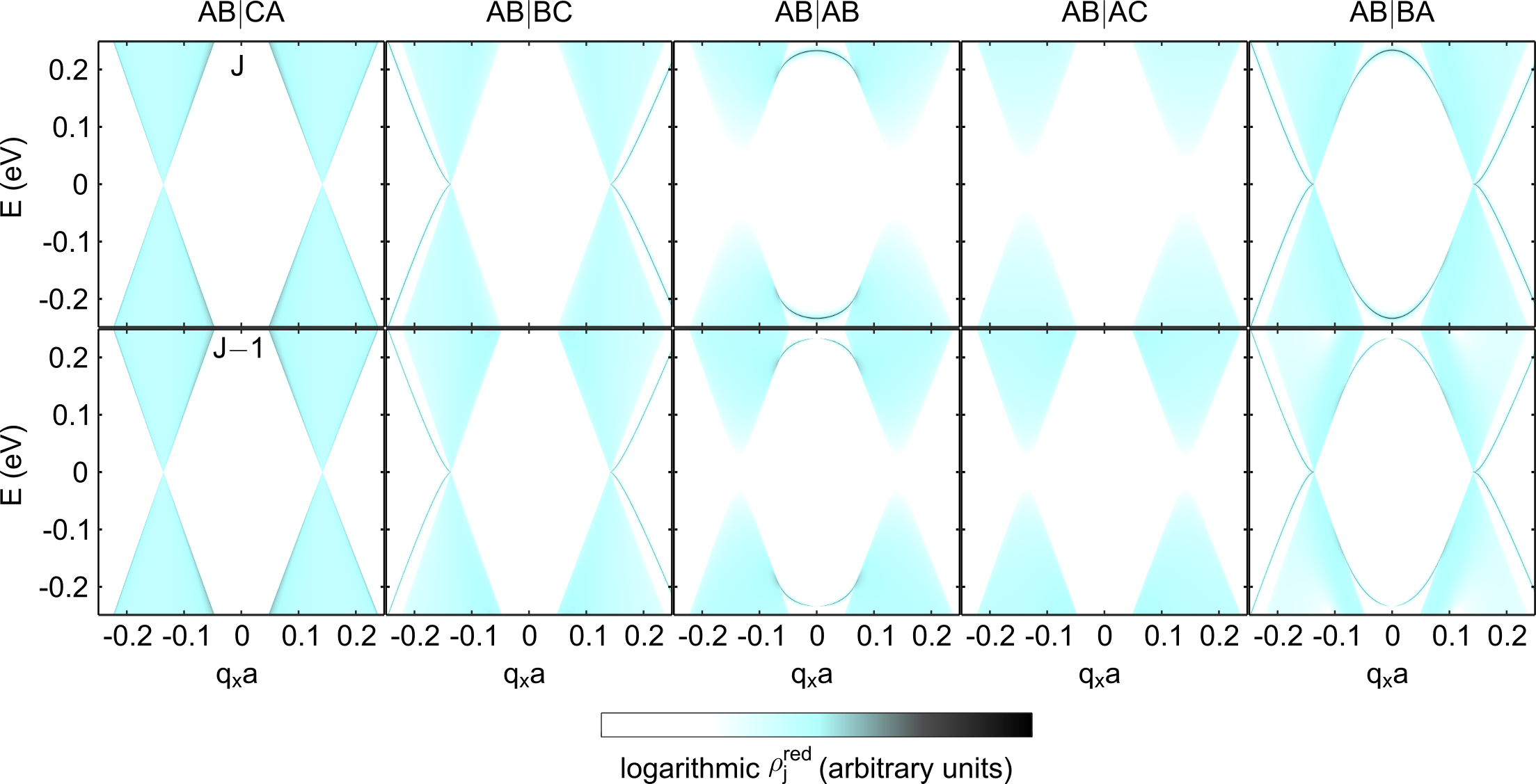}
\caption{\label{fig:redsite} 
\textbf{Low-energy local density of states.} The low-energy electronic density of states $\rho^{\mathrm{red}}_{j}$ on atoms of the red sublattice in layer $j=J$ (top row) and $j=J-1$ (bottom row) as a function of wave vector in the vicinity of the valley $\vect{K}=\left(\tfrac{4\pi}{3a},0\right)$. In these plots, we take $q_{y}=0$.
}
\end{figure}

\newpage

\section*{Supplementary Note 3: Analytical zero-energy states and dispersion relations}

Analytical expressions for the number of zero-energy states  (per spin per valley) $n^{\textrm{blue}}_{j}$ on atoms of the blue sublattice plotted in Fig.\@ 2 of the main paper were obtained via a low-energy expansion of the Green's function, and are given below. These are valid for the range of $\vect{q}$ such that $\gamma_{0}|f_{\vect{q}}|<\gamma_{1}$, where $f_{\vect{q}}=\exp{\left(i\tfrac{q_{y}a}{\sqrt{3}}\right)} - \exp{\left(-i\tfrac{q_{y}a}{2\sqrt{3}}\right)} \left(\cos\tfrac{q_{x}a}{2}+\sqrt{3}\sin\tfrac{q_{x}a}{2}\right)$;  $n^{\textrm{blue}}_{j} = 0$ otherwise. Layer $J$ is the layer directly to the left of the physical interface.

AB$|$BC:
\begin{equation}
n^{\textrm{blue}}_{J-m}(\vect{q}) =
\begin{cases}
\displaystyle\frac{1}{2}, & m=0, \\
\displaystyle\frac{\left(\gamma_{0}^{2}|f_{\vect{q}}|^{2}\right)^{m-1} \left(\gamma_{1}^{2}-\gamma_{0}^{2}|f_{\vect{q}}|^{2}\right)}{2\gamma_{1}^{2m}}, & m > 0.
\end{cases}
\end{equation}

AB$|$AB:
\begin{equation}
n^{\textrm{blue}}_{J-m}(\vect{q}) = \frac{\left(\gamma_{0}^{2}|f_{\vect{q}}|^{2}|\right)^{m} \left(\gamma_{1}^{2}-\gamma_{0}^{2}|f_{\vect{q}}|^{2}\right)}{\gamma_{1}^{2(m+1)}}, \ \ m \geq 0.
\end{equation}

AB$|$AC:
\begin{equation}
n^{\textrm{blue}}_{J-m}(\vect{q}) =
\begin{cases}
\displaystyle\frac{2\left(\gamma_{1}^{2}-\gamma_{0}^{2}|f_{\vect{q}}|^{2}\right)}{2\gamma_{1}^{2}-\gamma_{0}^{2}|f_{\vect{q}}|^{2}}, & m=0, \\
\displaystyle\frac{\left(\gamma_{0}^{2}|f_{\vect{q}}|^{2}|\right)^{m-1} \left(\gamma_{1}^{2}-\gamma_{0}^{2}|f_{\vect{q}}|^{2}\right)}{\gamma_{1}^{2(m-1)} \left(2\gamma_{1}^{2}-\gamma_{0}^{2}|f_{\vect{q}}|^{2}\right)}, & m > 0.
\end{cases}
\end{equation}

Dispersion relations for the non-topological dispersing states on atoms of the blue sublattice in layer $J$ were likewise determined from the poles of the Green's function. These are:

AB$|$BC:
\begin{equation}
|E| = \frac{\sqrt{2} \left(\gamma_{0}|f_{\vect{q}}|-\gamma_{1}\right)^{\frac{3}{2}}}{\left(2\gamma_{0}|f_{\vect{q}}|-\gamma_{1}\right)^{\frac{1}{2}}}.
\end{equation}

AB$|$AB:
\begin{equation}
\gamma_{0}^{2}|f_{\vect{q}}|^{2} = \frac{|E| \left( |E|^{2} + |E|\gamma_{1} - \gamma_{1}^{2} \right)}{|E|-\gamma_{1}}, \ |E| > \frac{\gamma_{1}}{2}.
\end{equation}

AB$|$BA:
\begin{subequations}
\begin{equation}
\gamma_{0}^{2}|f_{\vect{q}}|^{2} = \gamma_{1} \left(\gamma_{1} + |E|\right) + \sqrt{|E| \left(\gamma_{1} + |E|\right)^{3}}, \ |E| \geq 0,
\end{equation}
\begin{equation}
\gamma_{0}^{2}|f_{\vect{q}}|^{2} = \gamma_{1} \left(\gamma_{1} + |E|\right) - \sqrt{|E| \left(\gamma_{1} + |E|\right)^{3}}, \ |E| > \frac{\gamma_{1}}{3}.
\end{equation}
\end{subequations}

\pagebreak

\begin{figure}[h!]
\centering
\includegraphics[width=0.6\textwidth]{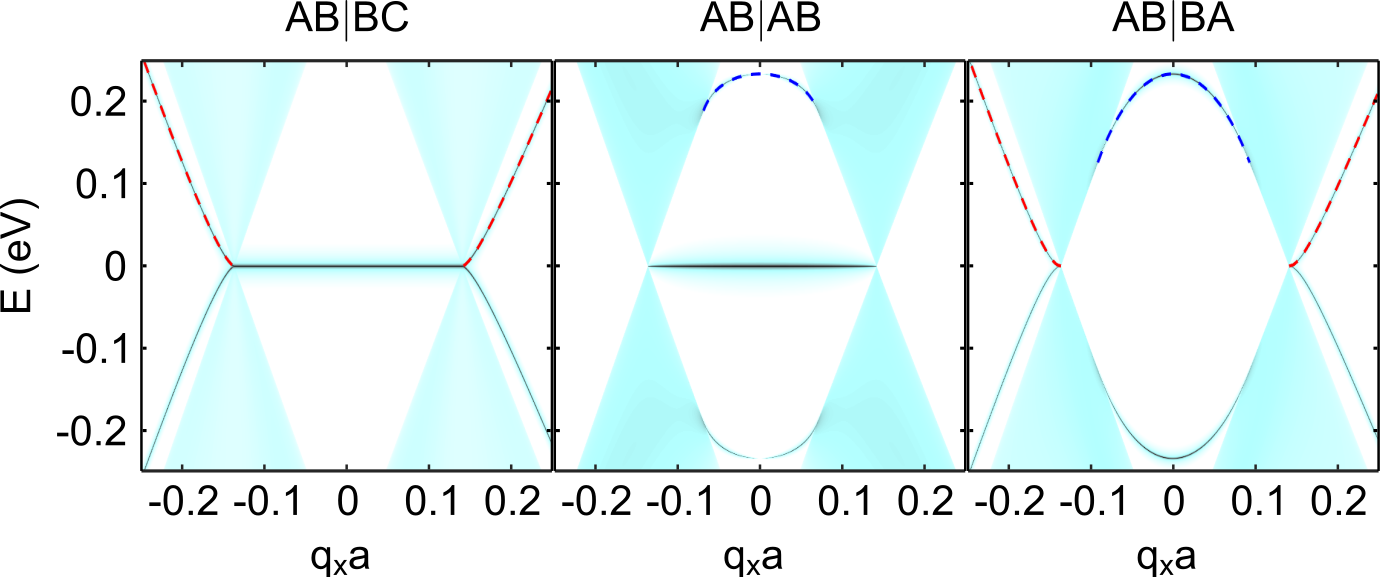}
\caption{\label{fig:dispersingstates} 
\textbf{Low-energy local density of states overlaid with analytic dispersion relations.} The low-energy electronic density of states $\rho^{\mathrm{blue}}_{j}$ on atoms of the blue sublattice in layer $J$ for the junctions that host dispersing states. The dispersion relations are shown with dashed lines; these are symmetric in energy about $E=0$ but only the dispersion for $E>0$ is highlighted, with colour differentiating between states possessing different dispersion relations. In these results we take $q_{y}=0$.
}
\end{figure}

\newpage

\section*{Supplementary Note 4: Sliding model}

Consider two layers of graphene that slide smoothly with respect to one another. To describe the sliding process we introduce a parameter $0 \leq \lambda \leq 3$.

\begin{figure*}[h!]
\centering
\includegraphics[width=0.85\textwidth]{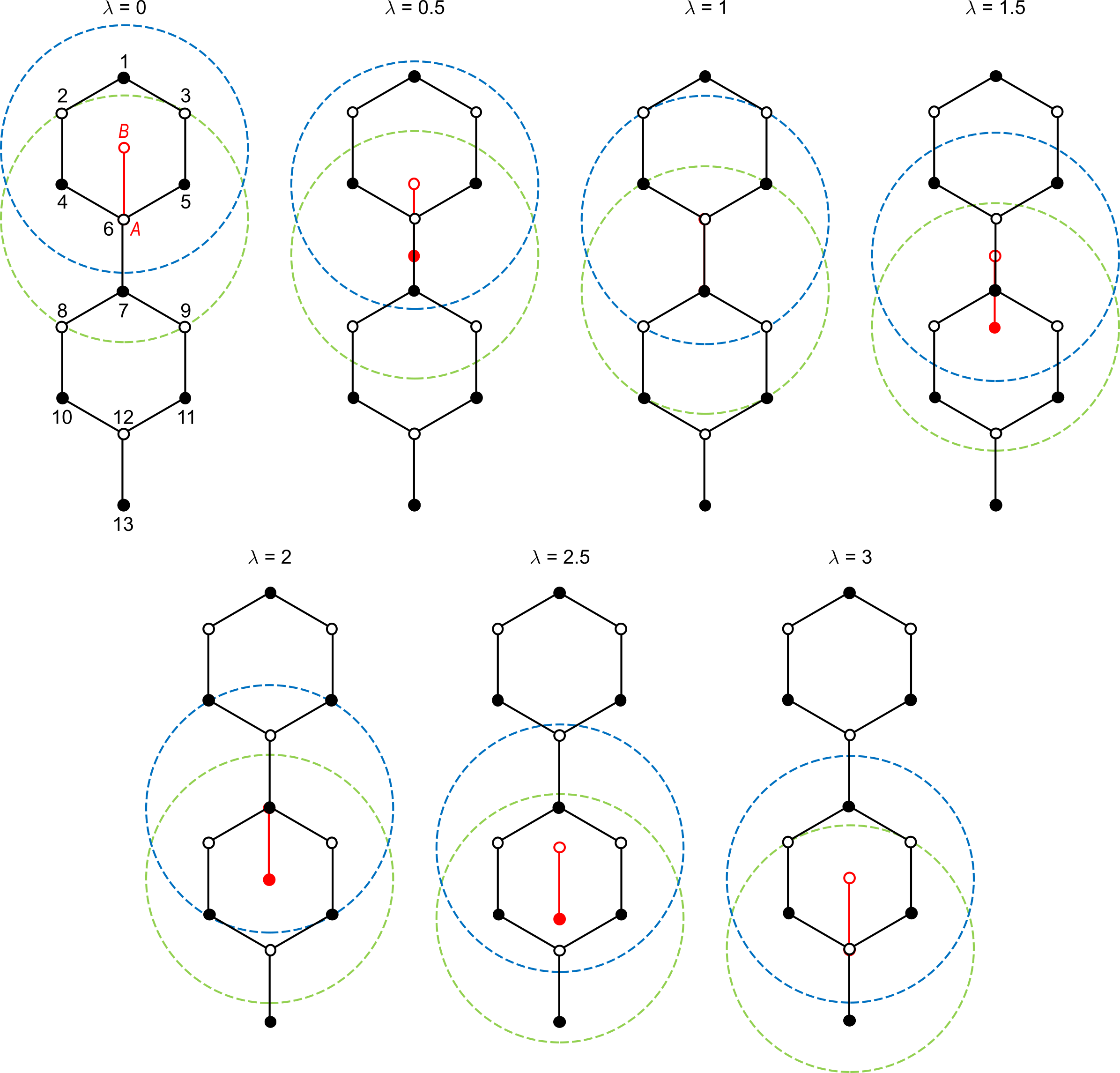}
\caption{\label{fig:sliding}
\textbf{Two layers of graphene that slide smoothly with respect to one another}. The positions of two atoms in layer $J$ are indicated in red, while the positions of the relevant atoms in layer $J+1$ are shown in black. Filled and empty circles differentiate between the two sublattices, which we refer to as $A$ and $B$ respectively. The layers are BA, AA, AB, and BA stacked at $\lambda = 0,1,2,3$ respectively. The green (blue) circle of radius $a$ contains all atoms in layer $J+1$ that couple to the $A$ ($B$) sublattice atom on layer $J$.
}
\end{figure*}

\pagebreak

As we neglect next-nearest neighbour interactions in the plane, we only need to consider 13 atoms on layer $J+1$ as these are the only ones that come within an in-plane distance of $|\vect{r}^{\mu,\mu'}|<a$ from either atom on layer $J$ during the sliding process, where indices denote the atom $\mu = A,B$ on layer $J$ and atom $\mu^{\prime}=$ 1, 2, 3, 4, 5, 6, 7, 8, 9, 10, 11, 12, 13 on layer $J+1$.

\begin{table}\label{table}
\caption{List of vectors $\vect{r}^{\mu,\mu^{\prime}}$, with $\mu = A,B$ and $\mu'=1,...,13$, used in our sliding model.}
\[
\begin{array}{|c|c|c|}
\hline
 & A & B \\
\hline
1 & \left( -\frac{a}{2}, \frac{a}{\sqrt{3}} \left( \lambda + 2 \right) \right) & \left( -\frac{a}{2}, \frac{a}{\sqrt{3}} \left( \lambda +1 \right) \right) \\
2 & \left( \frac{a}{2}, \frac{a}{\sqrt{3}} \left( \lambda + \frac{3}{2} \right) \right) & \left( \frac{a}{2}, \frac{a}{\sqrt{3}} \left( \lambda + \frac{1}{2} \right) \right) \\
3 & \left( 0, \frac{a}{\sqrt{3}} \left( \lambda + \frac{1}{2} \right) \right) & \left( 0, \frac{a}{\sqrt{3}} \left( \lambda - \frac{1}{2} \right) \right) \\
4 & \left( 0, \frac{a}{\sqrt{3}} \left( \lambda + \frac{1}{2} \right) \right) & \left( 0, \frac{a}{\sqrt{3}} \left( \lambda - \frac{1}{2} \right) \right) \\
5 & \left( -\frac{a}{2}, \frac{a}{\sqrt{3}} \left( \lambda + \frac{1}{2} \right) \right) & \left( -\frac{a}{2}, \frac{a}{\sqrt{3}} \left( \lambda - \frac{1}{2} \right) \right) \\
6 & \left( \frac{a}{2}, \frac{a}{\sqrt{3}} \lambda \right) & \left( \frac{a}{2}, \frac{a}{\sqrt{3}} \left( \lambda - 1 \right) \right) \\
7 & \left( -\frac{a}{2}, \frac{a}{\sqrt{3}} \left( \lambda - 1 \right) \right) & \left( -\frac{a}{2}, \frac{a}{\sqrt{3}} \left( \lambda - 2 \right) \right) \\
8 & \left( \frac{a}{2}, \frac{a}{\sqrt{3}} \left( \lambda - \frac{3}{2} \right) \right) & \left( \frac{a}{2}, \frac{a}{\sqrt{3}} \left( \lambda - \frac{5}{2} \right) \right) \\
9 & \left( 0, \frac{a}{\sqrt{3}} \left( \lambda - \frac{3}{2} \right) \right) & \left( 0, \frac{a}{\sqrt{3}} \left( \lambda - \frac{5}{2} \right) \right) \\
10 & \left( 0, \frac{a}{\sqrt{3}} \left( \lambda - \frac{5}{2} \right) \right) & \left( 0, \frac{a}{\sqrt{3}} \left( \lambda - \frac{7}{2} \right) \right) \\
11 & \left( 0, \frac{a}{\sqrt{3}} \left( \lambda - \frac{5}{2} \right) \right) & \left( 0, \frac{a}{\sqrt{3}} \left( \lambda - \frac{7}{2} \right) \right) \\
12 & \left( 0, \frac{a}{\sqrt{3}} \left( \lambda - 3 \right) \right) & \left( 0, \frac{a}{\sqrt{3}} \left( \lambda - 4 \right) \right) \\
13 & \left( 0, \frac{a}{\sqrt{3}} \left( \lambda - 4 \right) \right) & \left( 0, \frac{a}{\sqrt{3}} \left( \lambda - 5 \right) \right) \\
\hline
\end{array}
\]
\end{table}

The coupling matrix $\hat{V}$ between the layers has elements:
\begin{subequations}
\begin{equation}
\hat{V}_{A_{J},A_{J+1}} = \gamma^{A,4} + \gamma^{A,5} + \gamma^{A,7} + \gamma^{A,10} + \gamma^{A,11} + \gamma^{A,13},
\end{equation}
\begin{equation}
\hat{V}_{A_{J},B_{J+1}} = \gamma^{A,6} + \gamma^{A,8} + \gamma^{A,9} + \gamma^{A,12},
\end{equation}
\begin{equation}
\hat{V}_{B_{J},A_{J+1}} = \gamma^{B,1} + \gamma^{B,4} + \gamma^{B,5} + \gamma^{B,7} + \gamma^{B,10} + \gamma^{B,11},
\end{equation}
\begin{equation}
\hat{V}_{B_{J},B_{J+1}} = \gamma^{B,2} + \gamma^{B,3} + \gamma^{B,6} + \gamma^{B,8} + \gamma^{B,9} + \gamma^{B,12}.
\end{equation}
\end{subequations}

The in-plane vectors $\vect{r}^{\mu,\mu^{\prime}}$ connecting sites $\mu$ and $\mu^{\prime}$ for all atoms involved in this model are given in Supplementary Table 1.

The distance-dependent sigma bond integral $V_{pp\sigma}(d^{\mu,\mu'})$ and the angle $\theta^{\mu\mu'}$ between the out-of-plane direction and $\vect{r}^{\mu,\mu^{\prime}}$ are: 
\begin{subequations}
\begin{equation}
V_{pp\sigma} (d^{\mu,\mu^{\prime}}) = \frac{\gamma_1}{d_{\textrm{max}} - c} \left( d_{\textrm{max}} - d^{\mu,\mu^{\prime}} \right) \Theta \left( d_{\textrm{max}} - d^{\mu,\mu^{\prime}} \right),
\end{equation}
\begin{equation}
d^{\mu,\mu^{\prime}} = \sqrt{c^{2} + |\vect{r}^{\mu,\mu^{\prime}}|^{2}},
\end{equation}
\begin{equation}
\theta^{\mu,\mu^{\prime}} = \arctan \left( \frac{|\vect{r}^{\mu,\mu^{\prime}}|}{c} \right).
\end{equation}
\end{subequations}

\begin{figure*}[h!]
\centering
\includegraphics[width=0.4\textwidth]{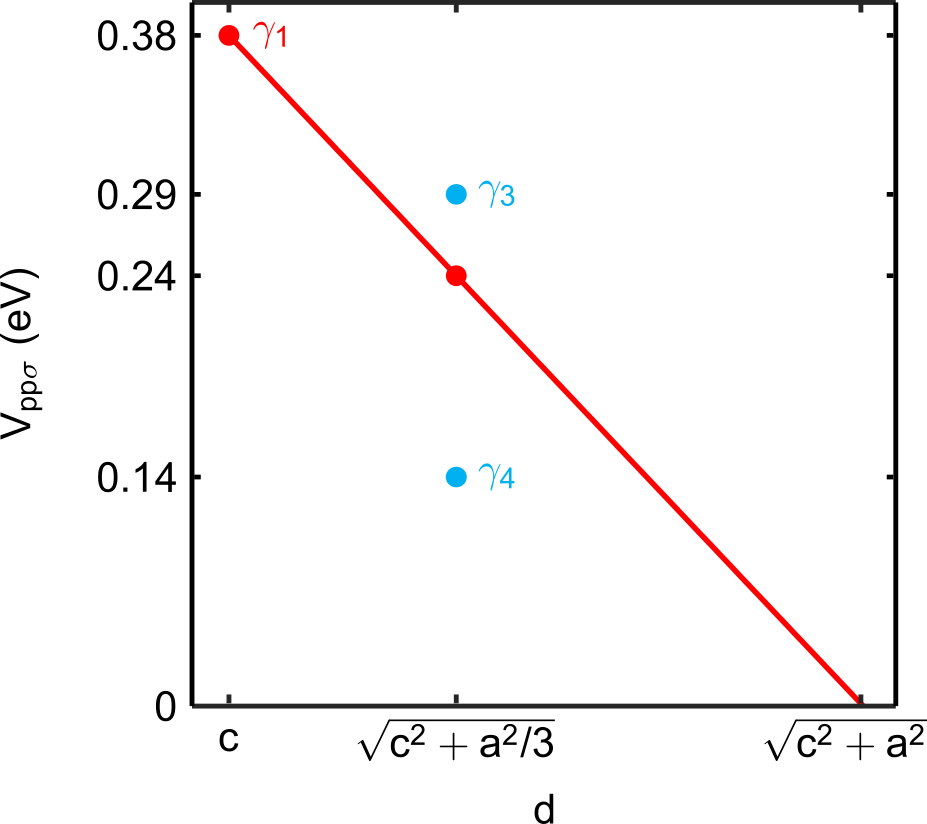}
\caption{\label{fig:hopping}
\textbf{Distance dependence of hopping parameter.} $V_{pp\sigma}$ used in the sliding calculations as a function of distance between atoms, taken to be zero beyond $\sqrt{c^{2}+a^{2}}$. The abrupt changes visible in Fig.\@ 3b of the main paper are due to of this sharp cut-off, and would not occur with a more smoothly decaying hopping. Here $\gamma_{3}$ and $\gamma_{4}$ are values of the interlayer coupling between atoms of differing and like sublattice in the bulk respectively. Their numerical values are taken from literature \cite{Shi2020, Slizovskiy2019, Kaladzhyan2021, Cea2022}.
}
\end{figure*}

\pagebreak
\newpage

\section*{Supplementary Note 5: Sliding results extended}

\begin{figure}[h!]
\centering
\includegraphics[width=0.75\textwidth]{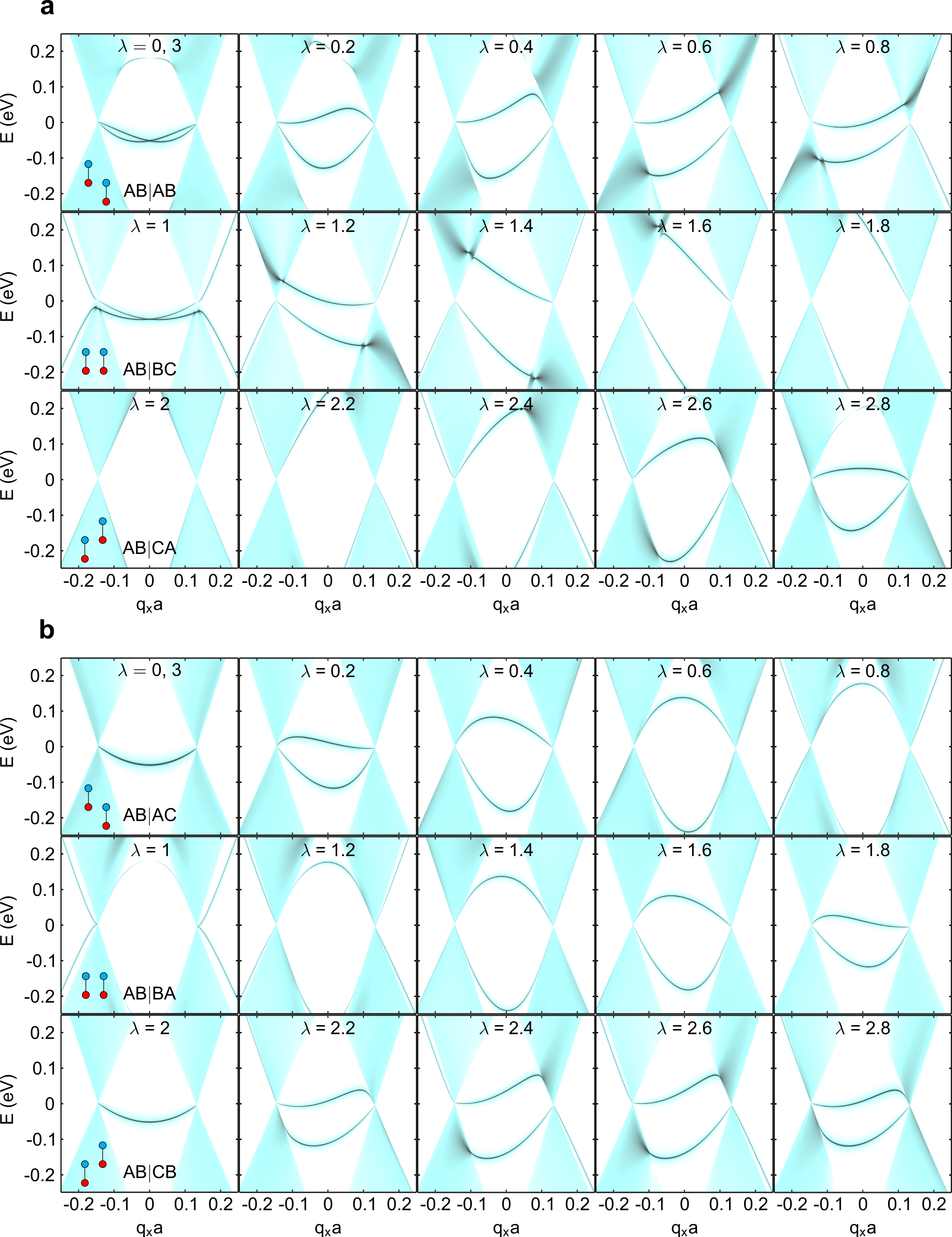}
\caption{\label{fig:slidingldosextended}
\textbf{Low-energy local density of states.} The low-energy electronic density of states on the blue sublattice in layer $J$ in the vicinity of the valley $\vect{K}$ for the \textbf{a}. AB$|$AB $\rightarrow$ AB$|$BC $\rightarrow$ AB$|$CA $\rightarrow$ AB$|$AB sliding junction, and the \textbf{b}. AB$|$AC $\rightarrow$ AB$|$BA $\rightarrow$ AB$|$CB $\rightarrow$ AB$|$AC sliding junction. Note that the AB$|$AC and AB$|$CB crystals are physically equivalent.
}
\end{figure}

\newpage

\section*{Supplementary Note 6: Rhombohedral trilayer on a half-crystal}

\begin{figure}[h!]
\centering
\includegraphics[width=0.8\textwidth]{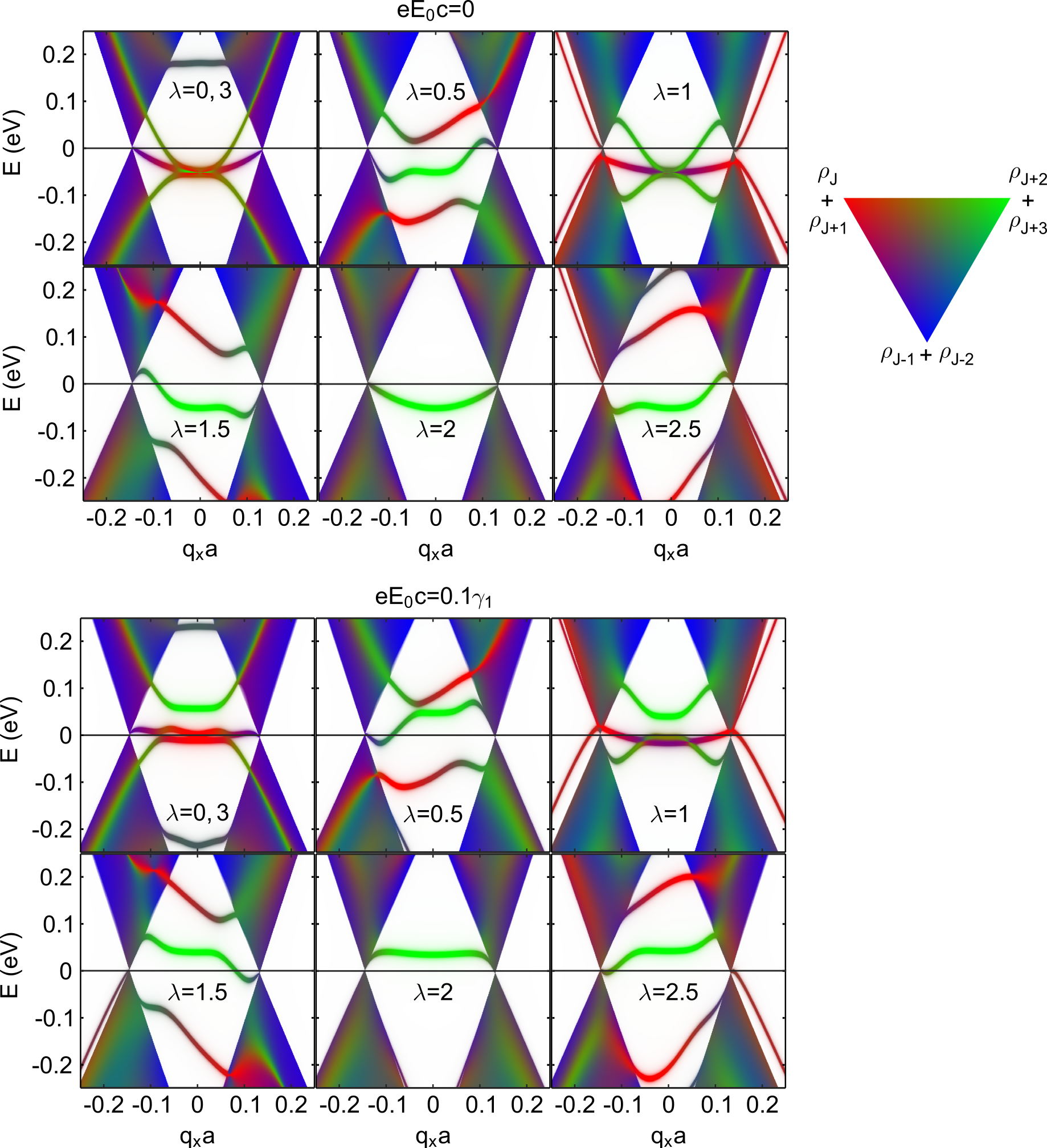}
\caption{\label{fig:gatingextended}\textbf{Junction states of a rhombohedral trilayer on a rhombohedral half-crystal.} In the top panel, no electric field is applied to the heterostructure. We use the same colour scheme as used in Fig.\@ 4 of the main text. In the bottom panel, for comparison, we show the results from the main text for an out-of-plane electric field $E_{0}$ such that $eE_{0}c=0.1\gamma_{1}$.
}
\end{figure}